\documentclass[twocolumn,showpacs]{revtex4}

\usepackage{graphicx}

\begin{document}

\title{Reweighting for Nonequilibrium Markov Processes Using \\
Sequential Importance Sampling Methods}

\author{Hwee Kuan Lee}
\affiliation{Department of Physics, Tokyo Metropolitan University,
Hachioji, Tokyo 192-0397, Japan}

\author{Yutaka Okabe}
\affiliation{Department of Physics, Tokyo Metropolitan University,
Hachioji, Tokyo 192-0397, Japan}

\date{\today}

\begin{abstract} 
We present a generic reweighting method for nonequilibrium Markov processes.
With nonequilibrium Monte Carlo simulations at a single temperature, one 
calculates the time evolution of physical quantities at different 
temperatures, which greatly saves the computational time. Using the dynamical
finite-size scaling analysis for the nonequilibrium relaxation, one can study
the dynamical properties of phase transitions together with the equilibrium 
ones. We demonstrate the procedure for the Ising model with the Metropolis 
algorithm, but the present formalism is general and can be applied to a variety
of systems as well as with different Monte Carlo update schemes.
\end{abstract}

\pacs{64.60.Ht, 75.40.Gb, 05.10.Ln}

\maketitle

The Monte Carlo simulation has served as a standard method to treat many body
problems in statistical physics~\cite{landau}. Metropolis 
{\it et al.}~\cite{metropolis} used the {\it importance sampling}, which 
generates the states with a probability proportional to the Boltzmann factor.
The {\it importance sampling} method can be considered as a reweighting 
technique in a sense that one samples a trial distribution and takes an 
average for the true distribution with reweighting. A more direct way of 
reweighting is found in the {\it histogram reweighting} method due 
to Ferrenberg and Swendsen~\cite{ferrenberg1}, where equilibrium thermal 
averages for a range of temperatures can be calculated from a single
simulation. This method greatly improved the efficiencies of Monte Carlo
simulations, however, the {\it histogram reweighting} technique has to be used
only to calculate equilibrium properties.

Recently, nonequilibrium relaxation method has been successfully applied to 
the study of critical phenomena~\cite{lou,zheng,ito,ozeki}. In the 
nonequilibrium relaxation method, simulations were performed for several 
temperatures; the critical temperature, the dynamical exponent and other 
quantities are estimated using the scaling behavior of nonequilibrium process.
If we combine the strength of nonequilibrium relaxation method with a 
reweighting technique, we can expect an effective method of simulation.

In this paper, we will present a generic reweighting method that is 
applicable to both equilibrium and nonequilibrium systems. With reweighting 
at nonequilibrium, only a simulation at a single temperature is required. Our
method of reweighting may also be applicable in a multitude of other 
nonequilibrium processes, 
such as the contact process \cite{henkel} and 
the driven diffusive systems \cite{schmittman}. 
%


Reweighting at nonequilibrium can be easily understood as follows. 
Consider every simulation up to the $t$th Monte Carlo step 
as a sequence of states (or path),
\begin{equation}
\vec{x}_t = (\sigma_1, \sigma_2, \cdots, \sigma_t)
\end{equation}
where $\sigma_t$ is the system configuration at time $t$. 
Hereafter, we refer to Monte Carlo step simply as the time 
of simulation.  Such path 
$\vec{x}_t$ can be generated using any Monte Carlo method at a temperature 
$T$. The objective is to calculate the relative probability of generating 
the same path $\vec{x}_t$ at another temperature $T'$.

Suppose many simulations were performed at an inverse temperature 
$\beta = 1/k_{\mbox{\tiny B}} T$ to obtain a set of paths 
$\vec{x}_t^j, j=1, \cdots, n$. (From now on, $\beta$ will be sometimes 
referred to as temperature).  Dynamical thermal average of 
some quantity $Q(t)$ can be obtained
by $\langle Q(t) \rangle_\beta = ( {1}/{n} ) \sum_{j=1}^n Q(\vec{x}_t^j)$. To
calculate the dynamical thermal average at another temperature $\beta'$, the 
measured quantity $Q(\vec{x}_t^j)$ has to be reweighted with a set of weights
$w_t^j$. For the same set of paths $\vec{x}_t^j$, the thermal average at 
$\beta'$ is obtained as
\begin{equation}
\langle Q(t) \rangle_{\beta'} = {\sum_{j=1}^n w_t^j Q(\vec{x}_t^j)}/
                                     {\sum_{j=1}^n w_t^j }
\label{eq:qw}
\end{equation}
Although not labeled explicitly in Eq.~(\ref{eq:qw}), the set of weights
$w_t^j$ depend on the simulation temperature $\beta$ and the new temperature 
$\beta'$. To calculate the weights, the following algorithm is implemented for
each path $\vec{x}_t^j, j = 1, \cdots, n$.
\begin{enumerate}
\item 
Assume that the Monte Carlo simulation is carried out at a temperature 
$\beta$, and a path $\vec{x}_t^j = (\sigma_1^j, \cdots, \sigma_t^j)$ for some
arbitrary time $t$ is obtained.
\item To go from time $t$, let $\sigma'^j$ be a trial configuration and
$T({\sigma '}^j | \sigma_t^j )$ be the probability to select this configuration.
If the trial move is accepted with the acceptance probability
$A_{\beta}({\sigma '}^j|\sigma_t^j)$, then the new configuration at $t+1$ 
becomes $\sigma_{t+1}^j = \sigma'^j$. If the move is not accepted,
$\sigma_{t+1}^j = \sigma_t^j$.
\item In terms of the transition probability 
$P_{\beta}(\sigma_{t+1}^j | \sigma_t^j)$, we may write as
$P_{\beta}(\sigma_{t+1}^j | \sigma_t^j) =
               T({\sigma '}^j|\sigma_t^j)
               A_{\beta}({\sigma '}^j|\sigma_t^j)$ 
	       if the move is accepted and,
$P_{\beta}(\sigma_{t+1}^j | \sigma_t^j) =
               T({\sigma '}^j|\sigma_t^j) [1-
	       A_{\beta}({\sigma '}^j|\sigma_t^j)]$ otherwise.
\item Then the appropriate weights $w_{t+1}^j$ can be obtained by
\begin{eqnarray}
w_{t+1}^j = \frac{P_{\beta'} (\sigma_{t+1}^j|\sigma_{t}^j)}
                 {P_{\beta } (\sigma_{t+1}^j|\sigma_{t}^j)} w_{t}^j & 
\mbox{with} & w_1^j = 1
\label{eq:w}
\end{eqnarray}
\item Repeat steps 2 to 4 until $t$ reaches some predetermined maximum 
simulation time.
\end{enumerate}
For each path $\vec{x}_t^j, j = 1,\cdots, n$, these steps are  repeated. Step
2 is simply the standard Monte Carlo procedure of acceptance and rejection.
Step 3 calculates the transition probability after the decision on acceptance
was made. Step 4 updates the appropriate weight. In this method,
essentially the usual Monte Carlo update is carried out, and at the same time
the weights were updated.

The proof of Eq.~(\ref{eq:w}) involves two ingredients, the Sequential 
Importance Sampling (SIS)~\cite{liu1,zhang,doucet} method 
and a generic Monte Carlo method. 
A brief description of SIS method will be discussed here. 
The SIS method 
assumes that the distribution for a multi-dimensional state 
$\vec{x}_t = (\sigma_1,\sigma_2,\cdots, \sigma_t)$ is known and let it be 
denoted by $\pi_t(\vec{x}_{t})$. Implementation of the SIS algorithm is as 
follows:
\begin{enumerate}
\item Draw $\sigma_{t+1}^j$ from a trial distribution 
$g_{t+1}(\sigma_{t+1}^j|\vec{x}_t)$ to form 
$\vec{x}_{t+1}^j = (\vec{x}_t^j,\sigma_{t+1}^j)$.
\item The importance weight at $t+1$ is given by
\begin{equation}
w_{t+1}^j = \frac{ \pi_{t+1}(\vec{x}_{t+1}^j) }
                 { \pi_{t  }(\vec{x}_{t  }^j) g_{t+1}(\sigma_{t+1}^j|\vec{x}_t^j)}
            w_{t}^j
\label{eq:sisw}
\end{equation}
\end{enumerate}
The key to understanding the reweighting method is to realize that in a Monte
Carlo simulation, the time sequence $\vec{x}_t^j$ is sampled from the true 
probability distribution of the path at some temperature $\beta$. The Monte 
Carlo method is Markovian and the probability of obtaining the path is given 
by
\begin{equation}
P_{\beta}( \vec{x}_t^j ) = P_{\beta}(\sigma_t^j | \sigma_{t-1}^j) \cdots 
P_{\beta}(\sigma_2^j | \sigma_1^j) 
P(\sigma_1^j)
\label{eq:markov}
\end{equation}
where $P_{\beta}(\sigma_t^j | \sigma_{t-1}^j)$ is the probability of getting 
a system configuration $\sigma_t^j$ at time $t$ given that the system 
configuration is $\sigma_{t-1}^j$ at an earlier time $t-1$. 
$P(\sigma_1^j)$ is the probability of choosing the initial configuration.
One may obtain the
probability distribution function, $P_{\beta'}(\vec{x}_t^j)$, of the path at 
another temperature $\beta'$ using $P_{\beta}(\vec{x}_t^j)$ as a trial 
distribution. The SIS method can then be used by defining the following 
quantities in Eq.~(\ref{eq:sisw}) as
\begin{eqnarray}
 \pi_{t+1}(\vec{x}_{t+1}^j) & \equiv &  P_{\beta'}(\vec{x}_{t+1}^j)\\
 \pi_{t}(\vec{x}_t^j) & \equiv &  P_{\beta'}(\vec{x}_t^j)    \\
 g_{t+1}(\sigma_{t+1}^j | \vec{x}_t^j ) & \equiv & P_{\beta}(\sigma_{t+1}^j | \vec{x}_t^j)
\end{eqnarray}
The requirement for the Monte Carlo process being Markovian implies
$P_{\beta}(\sigma_{t+1}^j|\vec{x}_t^j) = P_{\beta}(\sigma_{t+1}^j | \sigma_{t}^j)$, and
using Eq.~(\ref{eq:markov}), we obtain
\begin{equation}
w_{t+1}^j = \frac{ P_{\beta'}(\vec{x}_{t+1}^j)  }
{P_{\beta'}(\vec{x}_t^j) P_{\beta}(\sigma_{t+1}^j | \vec{x}_t^j)} w_t^j =
\frac{ P_{\beta'}(\sigma_{t+1}^j | \sigma_t^j)} 
{P_{\beta}(\sigma_{t+1}^j | \sigma_t^j)} w_t^j 
\end{equation}
which is as presented in Eq. (\ref{eq:w}). The initial condition in Eq. (\ref{eq:w})
is $w_1^j = 1$ because the initial system configuration is chosen from a distribution
independent of temperature.


To illustrate how the reweighting algorithm is implemented, we use the 
ferromagnetic Ising model on a square lattice. Its Hamiltonian is given by
\begin{equation}
{\cal H} = - \sum_{\langle k l \rangle} s_k s_l
\end{equation}
where the sum is over nearest neighbors and $s_k$ takes the values $\pm 1$. 
Periodic boundary conditions are used on a $L \times L$ lattice. We use the 
single spin flip update with the Metropolis acceptance rate, but other update
schemes are also applicable. 
The simulation is 
performed at $\beta$; the objective is to reweight it to $\beta'$. The system
configuration $\sigma$ is denoted by the states of all spins, 
$\sigma = \{ s_1, \cdots, s_N \} $, where $N$ is the total number of sites.
The initial system configuration at time $t=1$ is set to $s_k=+1$ for all 
$k=1,\cdots,N$. Hence $w_1^j = 1$ for $j=1,\cdots, n$; recall that $j$ is used
for indexing different paths. For each path 
$\vec{x}_t^j = (\sigma_1^j,\sigma_2^j,\cdots,\sigma_t^j)$,
\begin{enumerate}
\item Choose a spin and flip it with probability,
$A_{\beta}({\sigma'}^j|\sigma_{t}^j) = \mbox{min} [ 1, 
           \mbox{exp}(-\beta \Delta E) ]$.
Here, $\Delta E$ is the energy change due to the spin flip.
\item At this point, there are three possible outcomes,
  \begin{enumerate}
  \item If $\Delta E \leq 0$, the proposed spin flip is always accepted; and
  we obtain $w_{t+1}^j = w_{t}^j \times 1$.
  \item If $\Delta E > 0$ and the proposed spin flip is accepted, we obtain
  $w_{t+1}^j = w_{t}^j  \times \mbox{exp}(- (\beta' - \beta) \Delta E )$.
  \item If $\Delta E > 0$ and the proposed spin flip is rejected, we obtain
  $w_{t+1}^j = w_{t}^j \times {(1-\mbox{exp}(-\beta' \Delta E))}/
                {(1-\mbox{exp}(-\beta  \Delta E))}$.
  \end{enumerate}
\item Although the weights should be updated for every single spin flip move,
thermal quantities may be recorded only once per $N$ single spin flip steps,
for example, to save memory. Let $\tau$ be the time in unit of Monte Carlo 
steps per site (MCSS); weights $w_{\tau}^j$ and moments of magnetization 
${(m^\lambda)}_{\tau}^j, \lambda = 1,2,4$ (i.e. $m$, $m^2$ and $m^4$ moments)
are accumulated into the sums, $\sum_{j} w_{\tau}^j$,
$\sum_{j} {(m^\lambda)}_{\tau}^j$ and
$\sum_{j} {(m^\lambda)}_{\tau}^j w_{\tau}^j$.
\end{enumerate}
We carried out many simulations to obtain a set of paths $\vec{x}_t^j,
j = 1, \cdots, n$. Finally, the dynamical thermal averages for each time 
$\tau$, $\langle {m}(\tau)^\lambda \rangle_{\beta'}$, for example, 
can be calculated from the accumulated sums 
using Eq.~(\ref{eq:qw}).
In practice reweighting to many temperatures $\beta'_1,\beta'_2,\cdots$ 
were made at the same time in a single run.


\begin{figure}
\includegraphics[width=0.90\linewidth]{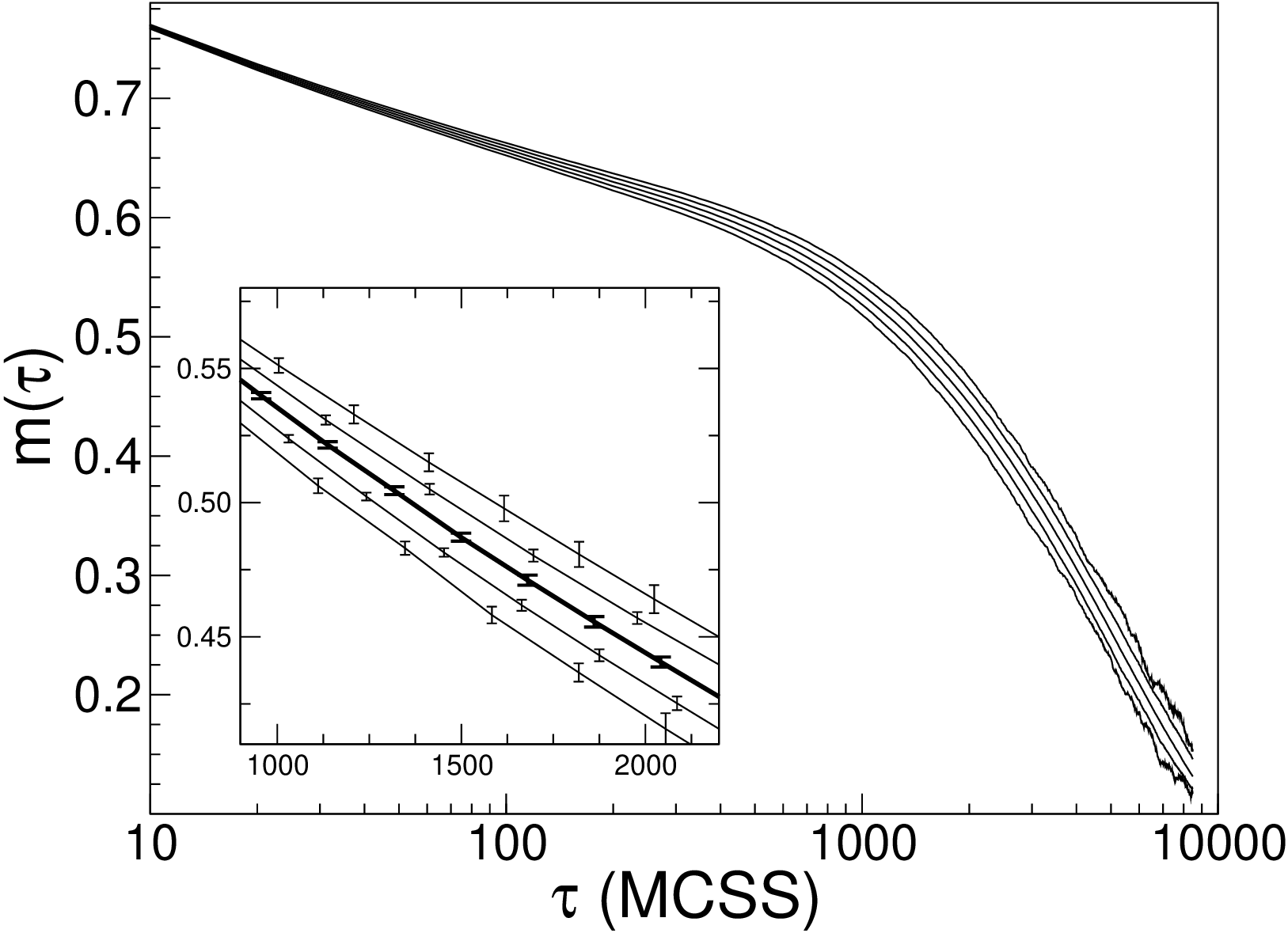}
\caption{Plot of magnetization for reweighted temperatures from 
simulations at a single temperature $T=2.270$ (bold line). From top to bottom,
$T=$2.268, 2.269, 2.270, 2.271, 2.272. System size is $L=64$. 
$\tau$ represents Monte Carlo time in units of MCSS.
Insert shows the region between $1000$ to $2000$ MCSS 
with representative error bars.
}
\label{fig:mt}
\end{figure}

Fig.~\ref{fig:mt} shows plots of the temporal evolution of magnetization for 
several reweighted temperatures from simulations at a single temperature 
$T=2.270$ (bold line) on the $L=64$ square lattice. These data were generated
with three independent runs to estimate error bars; for each run, averages 
were taken over $2.94\times 10^{5}$ samples. Effective reweighting temperature
range is about $\Delta T=0.002$ for $L=64$. Insert shows the region between 
$1000$ and $2000$ MCSS with representative error bars. 
Error bars become larger as reweighting is done at temperatures further away 
from the simulation temperature. 
Similar to the argument put forward by Ferrenberg and Swendsen 
\cite{ferrenberg1}, reweighting is effective when the distributions 
$P_{\beta}(\vec{x}_t^j)$ and $P_{\beta'}(\vec{x}_t^j)$ 
have sufficient overlaps.  An additional simulation, performed at 
$T=2.268$, for example, yields the results consistent with 
the reweighted ones within statistical errors. 
A more detailed comparison of actual and reweighted data 
will be discussed shortly.

\begin{figure}
\includegraphics[width=0.90\linewidth]{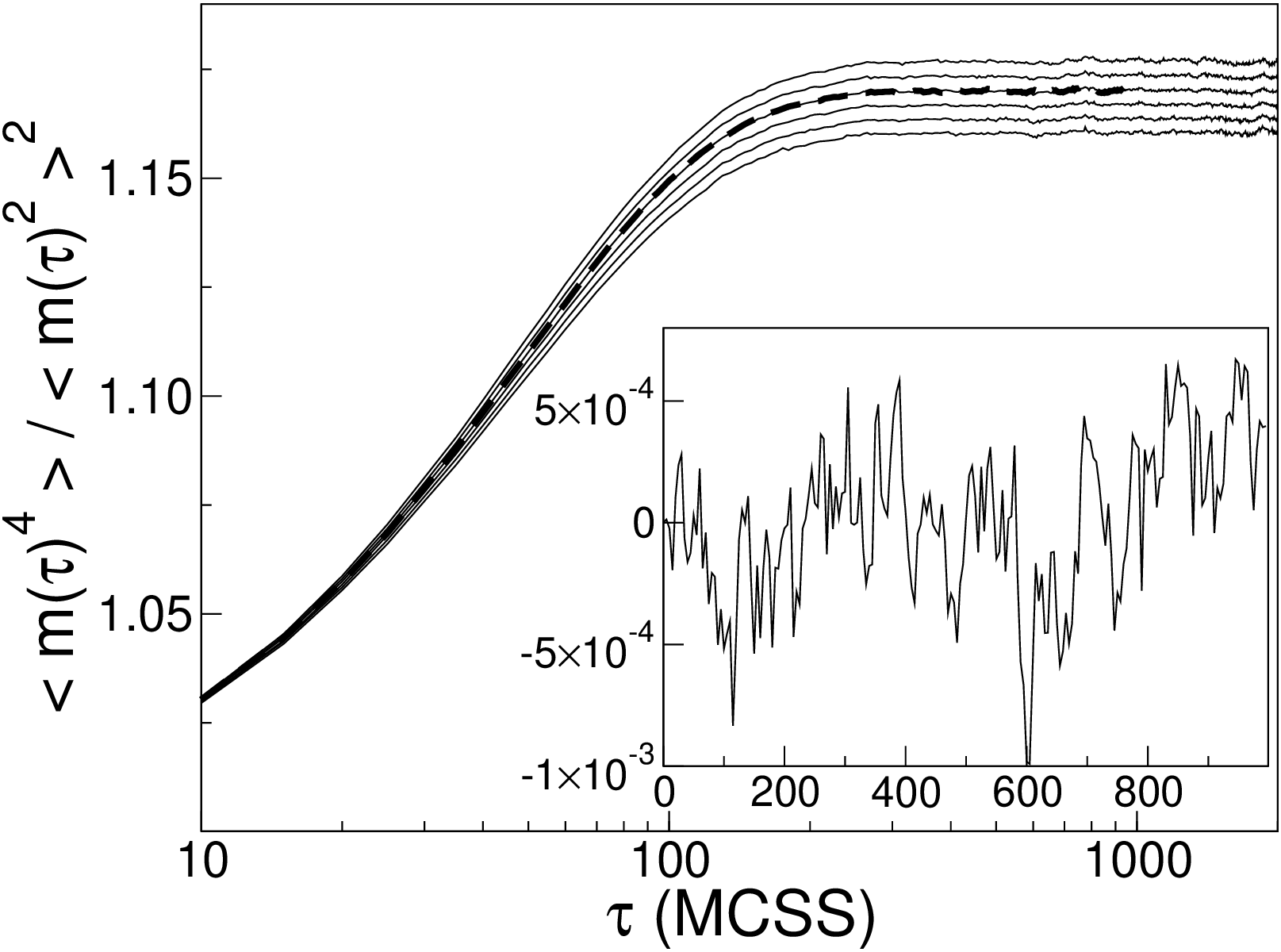}
\caption{Values of $\langle m(\tau)^4 \rangle / \langle m(\tau)^2 \rangle^2$ 
for reweighted temperatures from simulations at $T=2.270$. From top 
to bottom, $T = 2.272,$ $2.271$, $2.270$, $2.269$, $2,268$, $2.267$. To check
the range of reweighting, an additional simulation was performed at 
$T=2.268$ and reweighted to $T=2.270$. Dashed line shows
$\langle m(\tau)^4 \rangle / \langle m(\tau)^2 \rangle^2$ at $T=2.270$ reweighted 
from $T=2.268$. Insert shows that the difference between the dashed line and 
solid line is about $5 \times 10^{-4}$ while error bars on 
$\langle m(\tau)^4 \rangle / \langle m(\tau)^2 \rangle^2$ 
are of the order $10^{-3}$. The system size is $L=32$.
$\tau$ represents Monte Carlo time in units of MCSS.
} 
\label{fig:l32}
\end{figure}

The ratio of the moments  of the order parameter is used for the analysis of 
the phase transition~\cite{binder}. Fig.~\ref{fig:l32} shows plots of
$\langle m(\tau)^4 \rangle / \langle m(\tau)^2 \rangle^2$ 
for several reweighted temperatures from simulations at $T=2.270$. From top to
bottom, $T = 2.272,$ $2.271$, $2.270$, $2.269$, $2,268$, $2.267$. The lattice
size was $L=32$ and accuracy of reweighting was checked by performing an 
additional simulation at $T =2.268$. This simulation was then reweighted to 
$T=2.270$ and compared to the curve from the original simulation. The dashed 
line in Fig.~\ref{fig:l32} shows 
$\langle m(\tau)^4 \rangle / \langle m(\tau)^2 \rangle^2$ 
at $T=2.270$ reweighted from $T=2.268$. Insert shows that the difference 
between the dashed line and solid line is about $5 \times 10^{-4}$ while error
bars of $\langle m(\tau)^4 \rangle / \langle m(\tau)^2 \rangle^2$ 
are of the order $10^{-3}$. Averages were taken with $1.28 \times 10^6$ 
samples. The systematic errors due to reweighting are smaller than the 
statistical errors for this temperature shift.

\begin{figure}
\includegraphics[width=0.90\linewidth]{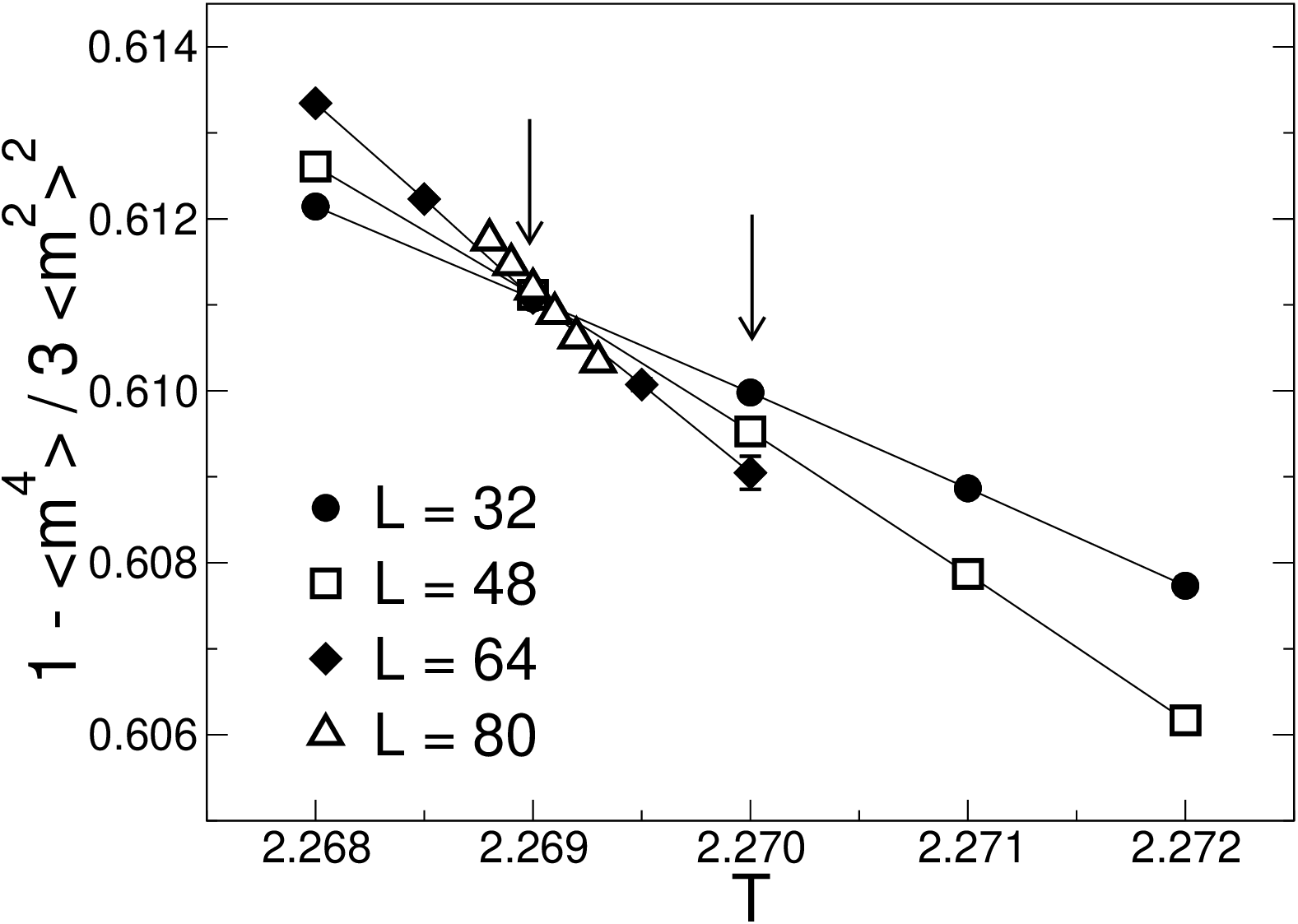}
\caption{Binder's cumulants for $L = 32$, $L=48$, $L=64$ and $L=80$. 
Simulation temperatures are indicated by arrows, $T=2.27$ for $L=32,48$, 
$T=2.269$ for $L=64,80$. Error bars when not shown are smaller than the size 
of the symbols.}
\label{fig:cross}
\end{figure}

Equilibrium properties can be calculated using reweighting by performing the 
simulation up to equilibrium and beyond, and then disregard data at 
nonequilibrium. In this domain, our reweighting method yields the same results
as the {\it histogram reweighting}~\cite{ferrenberg1} method.  
Fig.~\ref{fig:cross} shows Binder's cumulants~\cite{binder} 
for $L = 32$, $L=48$, $L=64$ and $L=80$. 
Simulations were performed at one temperature for each 
lattice size. The crossing of curves with different sizes yields the 
determination of the critical temperature $T_c$. In the present case, we 
confirm that the exact $T_c= 2.2691\cdots$ is well reproduced. Simulation 
temperatures are indicated by arrows; $T=2.27$ for $L=32,48$ and $T=2.269$ 
for $L=64,80$. The reweighting temperature range we used is $\Delta T=0.004$
for $L=32,48$, $\Delta T=0.002$ for $L=64$ and $\Delta T = 5 \times 10^{-4}$ 
for $L=80$. Error bars were generated with several independent simulations, 
and they are smaller than the size of the symbols. This suggests that the
effective reweighting temperature range is actually larger than the 
temperature range we used in our calculations. For each independent run, 
averages were taken with $1.28 \times 10^6$ samples for $L=32$,
$2.59 \times 10^5$ samples for $L=48$, $1.28 \times 10^5$ samples for
$L=64$ and $4 \times 10^4$ samples for $L=80$.

\begin{figure}
\includegraphics[width=0.90\linewidth]{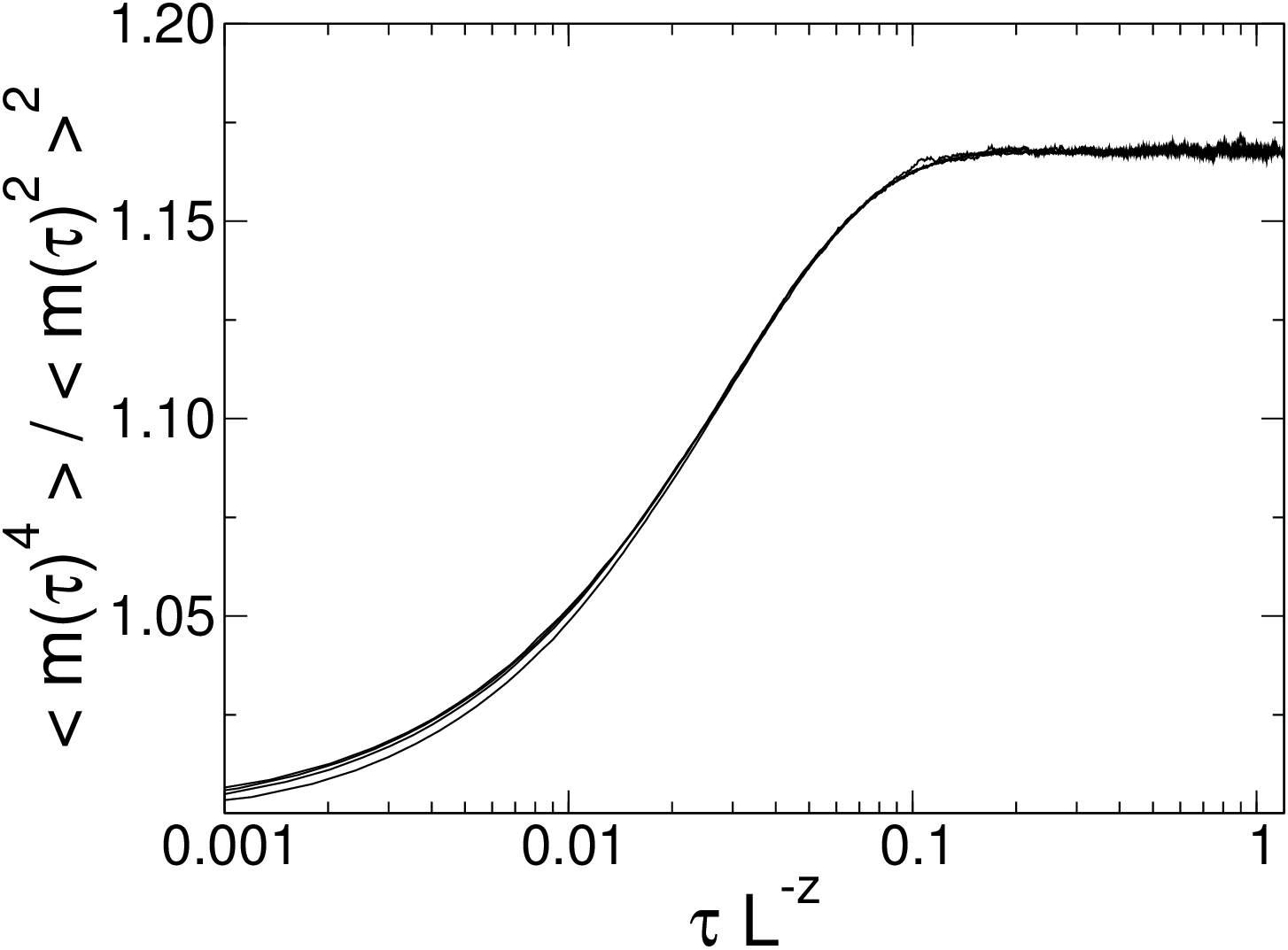}
\caption{Finite-size scaling plot of 
$\langle m(\tau)^4 \rangle / \langle m(\tau)^2 \rangle^2$ for $L=32, 48, 64$ 
and $80$ at $T_c = 2.2691\cdots$. The dynamical exponent is chosen as
$z=2.143$ for this plot.}
\label{fig:collapsed}
\end{figure}

The dynamical finite-size scaling~\cite{suzuki} can be used in combination
with nonequilibrium relaxation~\cite{kikuchi,goldschmidt}. Making use of the 
scaling form~\cite{okabe},
\begin{equation}
\langle m(\tau)^4 \rangle / \langle m(\tau)^2 \rangle^2 = g( \tau L^{-z})
\end{equation}
the dynamical exponent $z$ can be estimated by plotting 
$\langle m(\tau)^4 \rangle / \langle m(\tau)^2 \rangle^2$ 
versus $\tau L^{-z}$ at the known $T_c = 2.2691\cdots$ value 
(Fig.~\ref{fig:collapsed}). The estimate of $z$ is obtained when curves of 
different sizes are collapsed into a single curve. The best fit for data 
collapse is obtained by choosing $z=2.143 \pm 0.063$ for all the data shown 
in Fig.~\ref{fig:collapsed}. For the curves of $L=32$ 
and $48$, the best fit occurs at $z=2.11 \pm 0.02$. The best fit is obtained 
at $z=2.138\pm 0.039$ for $L=48$ and $64$, and at $z=2.143 \pm 0.063$ for 
$L=64$ and $80$. 
We have given the error bars on $z$ from the variance of 
the estimate of $z$ by a set of independent runs. 
Simulations were performed at relatively small system sizes 
compared to more extensive simulations~\cite{okabe,ito1}. Considering the 
effect of corrections to scaling, our value of $z$ is, within error bars, 
approaching the values that were previously reported 
$z=2.16 \sim 2.17$~\cite{okabe,ito1,nightingale}.

To summarize, we have presented a reweighting method for nonequilibrium Markov
processes. We have shown the basis of this method starting from the 
formulation of the SIS method. As a demonstration, we have used the Ising 
model on a square lattice. With the nonequilibrium simulation at a single 
temperature, we can determine the critical temperature and critical exponents
using finite-size scaling. The nonequilibrium simulation for large enough 
systems~\cite{ito,ozeki} is one way to extract dynamical and also static 
properties without worrying about finite-size effects. The systematic analysis
of finite-size effects, the finite-size
scaling~\cite{suzuki,kikuchi,goldschmidt}, is another way of studying 
nonequilibrium relaxation. The nonequilibrium reweighting is useful when it is
combined with the latter approach. We should note that this method is neither
restricted to single spin flip updates nor the Ising model. This method is
applicable, in principle, whenever the ratio of probabilities 
$P_{\beta'}(\sigma_{t+1}^j|\sigma_{t}^j)/P_\beta(\sigma_{t+1}^j|\sigma_{t}^j)$ 
can be calculated numerically. It should be applicable with other Monte Carlo
update schemes such as 
the cluster update schemes~\cite{wang,wolff} 
as well as the $N$-fold way method~\cite{bortz}. 
We should mention that Dickman~\cite{dickman} proposed a similar reweighting 
method for the application to the contact process. However, our formalism is 
more extensive and general.

This work was supported by a Grant-in-Aid for Scientific Research from
the Japan Society for the Promotion of Science. The computation of this work
has been done using computer facilities of the Supercomputer Center, Institute
of Solid State Physics, University of Tokyo.


\end{document}